# Digital Passport and Visa Asset Management Using Private and Permissioned Blockchain

Keenu Chandra
*Department of Computer Science*
*Nitte Meenakshi Institute of Technology*
Bangalore, India
1nt17cs085.keenu@nmit.ac.in

Maroof Mushtaq
*Department of Computer Science*
*Nitte Meenakshi Institute of Technology*
Bangalore, India
1nt17cs102.maroof@nmit.ac.in

Dr. Nalini N.
*Professor, Dept. of CS&E NMIT*
Bangalore, India
nalini.n@nmit.ac.in

*Abstract*—Blockchain is currently one of the fastest-growing technologies in the field of Computer Science. It has found a prevalent use in financial applications like cryptocurrency, for example, Bitcoin and Ethereum. They have been able to bring an unforeseen disruption in the field of finance. However, permissionless Blockchains like these have some downsides, namely the computation cost of the Proof of Work algorithm, maximum allowed size for a block, decrease in intelligibility with the increase of the number of blocks in the chain, domination of nodes with higher computing power as miners and validators. These factors have restricted the adoption of permissionless blockchain technology outside the field of finance, such as in medical or legal fields. This paper proposes a solution to these problems using a permissioned blockchain. It does not require a computationally expensive consensus mechanism as permissioned chains call for trust between participating organizations which is achieved via exclusive invitations. We have utilized a third-party orderer to maintain the trust between organizations.

## Introduction

The traditional asset management system for physical VISA and passports has numerous limitations that arise essentially due to the asset control and storage dynamics. It is still paper-based despite all the advancements in digital technology. Traditionally assets are regulated by a single organization and stored at a central database, which on its own is susceptible to human error, forgery, or theft. The process of applying for a passport or a VISA is quite complicated and involves manual labour and time, and is vastly inefficient. This process can also get overly repetitive in case multiple VISA is required for the same country [1]. There have also been numerous cases of passport frauds[2] that may lead to a bigger disaster depending on the fraudster's intentions. It is a threat to national and international trust and security. Apart from that, the problem of passport confiscation of immigrant labourers is extremely common which leads to their exploitation and other human rights issues[3][4]. There can also be honest accidents of people losing their physical passport or VISA that can land them into trouble abroad[5]. Reissuing passports in such cases is a time and resource-consuming, repetitive, and tedious process. Many countries are trying to find a more efficient solution to this problem[6].

In this paper, we propose a solution using Hyperledger, a permissioned blockchain, that will allow simplifying passport and VISA issuing, verification, and validation processes to make it streamlined and seamless. It creates a distributed ledger to store the passport data of all the citizens of a country along with their visa application record and travel history. We have decided to utilize CouchDB, an on-chain database, to store the data that ensures security and immutability, which is highly essential for sensitive data like this. There are four main components in the proposed system: a passport issuing authority, a visa office, a third-party orderer that acts as the channel admin in the framework that can read, write, and update data to and from the chain, and a web-based interface for users (citizens of the country) where they can apply for passport or VISA, check the status of the application, and produce a trustworthy digital copy of passport or VISA wherever and whenever required. The final product automates many processes and makes the whole process much simpler for all the humans involved.

We will first provide background information about permissionless blockchain architecture (Bitcoin and Ethereum) and permissioned blockchain architecture (Hyperledger) in section II. In section III, we will discuss previous attempts at digitizing passports and other works related to our project. In section IV, we will describe our proposed solution, its implementation, and the architectural model in detail. We provide the summary of our work and the scope of future work in section V of this paper.

## Background

Blockchain is a distributed and immutable ledger that is shared among all the participants of a business network but does not have any central authority[7]. It allows the network to record transactions and track assets. An asset can be any intangible or tangible object that holds value. Blockchain maintains a list of records of all the assets that are called blocks. These blocks are encrypted using complex cryptographic functions to make them immune to tampering. The three key elements of blockchain are distributed ledger technology, immutable records, and smart contracts[8]. Together they allow all the participants of the network to access the distributed ledger and perform authorised actions on it. All the transactions are recorded only once and no participant can change or tamper with the transaction after it has been recorded in the shared ledger. The smart contract is a set of rules that is stored on the chain and is responsible for defining conditions for the allowed transactions[8].

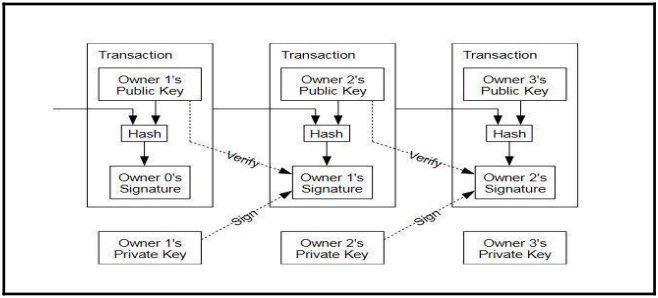

*Figure 1. Transaction in Bitcoin*

*Permissionless Blockchain*

Permissionless blockchain is essentially a public blockchain framework. It is a shared network that does not impose any restriction on participation[9]. Anyone from around the globe can join and become a validator in such blockchain networks. These networks are truly decentralised and have maximum security and transparency. All the popular blockchains that we know of today such as Bitcoin, Ethereum, and Litecoin are permissionless or public blockchain. Complete copies of the ledger in these chains are copied and stored around the globe, granting it extreme immunity to hacking and tampering. The participants can remain anonymous as there is no need to provide identity to gain access to the system[10].

Bitcoin was developed to invent an electronic payment that is based on cryptographic proof instead of just vesting trust third parties to process the payments. In Bitcoin, the owner of an electronic coin can transfer it to someone digitally by signing a hash of the previous transaction and the public key of the next owner. The second party will then verify the signatures to complete the transaction[11]. A trusted central authority is established to check transactions for double spending.

Ethereum aims to create an alternative protocol for building decentralized applications. It provides a blockchain with built-in Turing-complete programming language. It allows anyone to write their own smart contracts[12]. Although both the technologies work well for their applied field, using a permissionless blockchain is not adequate for our project.

*Permissioned Blockchain*

Contrary to permissionless blockchains, a permissioned blockchain requires explicit approval from the network to access the chain and ledger. These can be thought of as private networks that act as closed ecosystems and allow only those to view or validate the transactions that are approved by the central authority[10]. They are much less transparent as compared to permissionless chains, hence are suitable for instances where sensitive data is being handled. They are useful for institutions that want complete control over their data, such as banks and private companies.

Examples of permissioned blockchain are Ripple and Hyperledger.

Hyperledger is a joint initiative by the Linux Foundation and IBM. It started in 2015 with the cooperation of many different companies along with the Linux Foundation. It has many frameworks like Hyperledger Burrow, Hyperledger Sawtooth, Hyperledger Quilt, and Hyperledger Fabric. It is designed to be modular, highly secure interoperable cryptocurrency agnostic and complete with APIs[17][19]. For our project, we are going to use Hyperledger Fabric. It is a permissioned blockchain where all the nodes have an identity and there exists a Membership Service Provider (MSP) that uses a public key to issue cryptographic certificates to the participants. It uses a consensus mechanism that reduces computational complexity many fold when compared to proof of work mechanism. Fabric also has smart contracts, Chaincode, and the ability to create channels. The architecture of Hyperledger Fabric is modular in nature and allows the user flexibility in consensus mechanism, MSP, certificate authority, and interface SDK. It can be used to build an efficient, secure, and flexible blockchain platform for enterprise use[17-20].

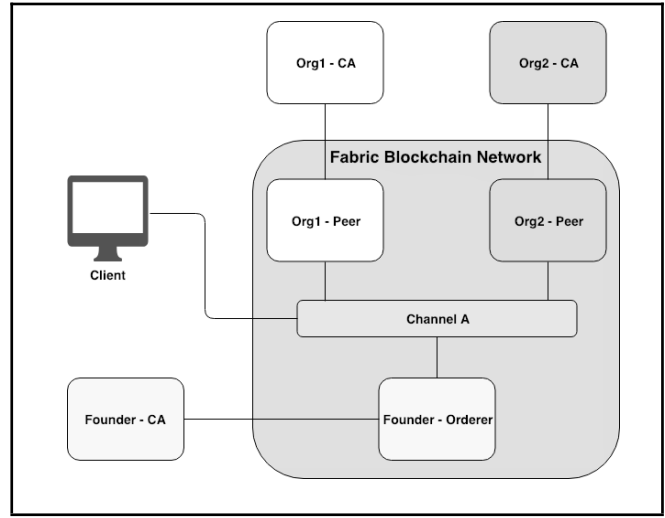

*Figure 3. Core structure of Hyperledger Fabric*

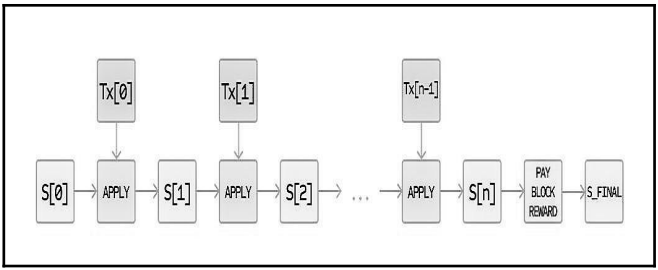

*Figure 2. Mining process in Ethereum*

RELATED WORKS

The "Known Traveler Digital Identity" [13] was initiated at the World Economic Forum (WEF) aiming to unlock the potential of digital identity for safe and seamless travel. This system exploits different innovations such as blockchain, IoT and machine learning technologies to build a prototype. The main goal of this project is to minimize trans-border and cybersecurity threats. The Known Traveler Digital Identity system allows travellers to use a mobile application to store and share their travel documents with authorities before the screening process. All the data related to the international passengers will be stored on the blockchain.

This system uses Consortium Blockchain, a semi-private blockchain that is controlled by a trusted group of organisations but any organisation can use it as a public blockchain. VeriDoc global [14] is a digital passport storage and verification system. This software solution tackles fake documentation problems and verifies documents. When a user or a citizen gets a passport from the Government Passport Office, VeriDoc scans the passport, puts a QR code on it and converts it into a PDF format. The PDF is stored on secure servers and hash is generated for the pdf URL and placed into the blockchain. VeriDoc maintains the off-chain database to match the pdf URL to the hash. Users or citizens can simply log in to the server, provide their hash ID and the passport, in the form of a PDF, will be retrieved. Knowledge Media Institute (KMI) which is a part of Open University UK (OU) has built an asset management and verification system to share and verify course completion certificates and badges. KMI's system uses permissionless blockchain technology, like Ethereum. KMI is collaborating with other institutes to build a universal system, which will make it easy for students to share their certificates and badges among different institutes seamlessly and securely[15]. Another similar solution for educational certificate verification has also been proposed by O. S. Saleh et al [16].

SYSTEM IMPLEMENTATION

In this section, we will discuss our proposed system in detail. Our proposed system provides a safe and secure alternative to physical and digital passport and VISA documents. This system implements an e-Identification system to verify the users across passport and VISA offices. We will describe the logic of how the Passport office can verify and issue a passport to a citizen and how citizens can conveniently apply for a VISA. We will also discuss different parts of Smart Contract that form our business logic.

*System Structure*

Like any other system based on Hyperledger Fabric technology, our proposed system also includes components such as organisations, orderer, channel and ledger.

*Organisations:*

These are the companies that define the rules and regulations to be implemented on the blockchain. In our case, these companies (or organisations) are the Indian Passport office and VISA offices of other nations. Every organisation in Hyperledger Fabric has a CA (Certificate Authorities), peer(s), MSP (Membership Service Provider), and a copy of the shared ledger.

*Certificate Authorities and MSP:*

CAs are the nodes that issue certificates of identity to all the peers or users. The abstraction of these certificates or identities is provided by the MSP definitions. MSP provides an abstraction of membership operations. It defines all the protocols and mechanisms for generating, issuing and verifying the certificates as well as user authentication.

*Peer(s):*

A peer is an essential component of the organisation that hosts Chaincode and ledgers. A peer is the user of the organisation. Usually a single organisation is composed of multiple peers. MSP assigns each peer their permission to interact with the blockchain and defines which operations are allowed. A peer can have multiple chaincodes installed or can host multiple ledgers.

*Ledger:*

A ledger is the database system of a blockchain. It stores transactions in the form of key-value pairs. Hyperledger Fabric ledger has two parts - World state and Blockchain. World state is a Database that stores the current values of a ledger state. Blockchain is a transaction log that stores all the changes that have been made to the world state.

*Smart Contract:*

A Smart Contract (also known as "Chaincode" in Hyperledger technology) contains the executable business logic. It defines the rules and permissions on how the businesses interact with each other, as well as the blockchain network. It is a piece of code written using a computer programming language, like JavaScript, Go, or Java. This code allows the users to perform operations like read and write on the ledger. Users interact with it using SDK's. It also defines the structure of objects (or assets) that can be stored, retrieved, or updated on the blockchain. Chaincode interface has three major components: a Deploy method to install the chaincode on a peer, an Invoke method to update or write a smart contract, and a Query method to read or query the current state of the ledger.

Definition of assets in our Smart Contract is shown below:

*User asset:*

abstract asset User identified by userId {

o String userId

o String name

o String email

o Integer phoneNumber

o String address

o Integer aadhaarNumber

}

*Passport asset:*

abstract asset Passport identified by passportId {

o String passportId

o String name

o String email

o Integer phoneNumber

o String address

o Integer aadhaarNumber

o DateTime issueDate

}

*VISA asset:*

abstract asset VISA identified by visaId {

o String visaId

o String country

o String visaType

o String passportId

o String name

o String email

o String address

o Integer aadhaarNumber

o DateTime visaIssueDate

o DateTime visaExpireDate

}

Smart Contracts are essential in maintaining the permissions and the operations allowed to a user or a peer. These ensure that the data is read or queried by the permissible users only and data written to the ledger is entered by the authorized user with permission to write on the ledger.

Policies:  
Readers:  
   Type: Signature  
   Rule: "OR('OrdererMSP.member')"  
Writers:  
   Type: Signature  
   Rule: "OR('OrdererMSP.member')"  
Admins:  
   Type: Signature  
   Rule: "OR('OrdererMSP.admin')"

In order to ensure that the data present on the blockchain is valid and verified by the authorized users, we have used a strong permission logic. For example, permission logic for passport issuing authority to view user applications, verify users, issue a passport and write passport details on the ledger. Similarly permission logic for VISA issuing authority to view only valid users, verify users with the passport office of the users country, issuing a VISA and writing VISA details to the ledger.

*Algorithms*

The algorithms used in this paper are as follows:

### ALGORITHM 1- Issuing Passport

Step 1: Start  
Step 2: Passport issuing agent logs in using correct credentials.  
Step 3: These credentials are matched with the credentials stored on the blockchain.  
    Step 4: If the credentials are correct:  
        Step 5: A list of Passport applications are queried from the blockchain and displayed to the agent.  
        Step 6: The agent reviews the applications  
        Step 7: If the application is accepted:  
            Step 8: Agent write the passport details on the blockchain.  
        Step 9: Else, Application request and data is deleted from the blockchain.  
    Step 10: Repeat Step 4 for other pending requests.  
    Step 11: Else, the page returns to the login page.  
Step 12: Stop.

### ALGORITHM 2- Issuing VISA

Step 1: Start  
Step 2: VISA issuing agent logs in using correct credentials.  
Step 3: These credentials are matched with the credentials stored on the blockchain.  
    Step 4: If the credentials are correct:  
        Step 5: The list of VISA applications are queried from the blockchain and displayed to the agent.  
        Step 6: The agent reviews and verifies the application with the passport office of the user's country.  
        Step 7: If the application is accepted:  
            Step 8: Agent write the VISA details on the blockchain.  
        Step 9: Else, Application request and data is deleted from the blockchain.

    Step 10: Repeat Step 4 for other pending requests.  
    Step 11: Else, the page returns to the login page.  
Step 12: Stop.

### ALGORITHM 3- Querying Passport and VISA

Step 1: Start  
Step 2: User logs in using correct credentials.  
Step 3: These credentials are matched with the credentials stored on the blockchain.  
    Step 4: If the credentials are correct:  
        Step 5: The corresponding Passport is queried from the blackchain and displayed to the user.  
        Step 6: The user can also query the approved VISA's.  
    Step 11: Else, the page returns to the login page.  
Step 12: Stop.

### WORKFLOW

First, a user has to apply for a passport, if they don't one already. Passport registrations don't need any credentials as fthe user is new to the system. After filling and submitting an application form online, this application form is written on the blockchain. Passport issuing authorities, which need credentials to log in to the system, query all the pending passport application forms from the blockchain database. User verification is done based on the information submitted by the user through the application. If the user verification is successful then the passport issuing agent writes the user data along with the passport identification on the blockchain. If the user verification fails, the application request is deleted from the blockchain. If a passport is issued successfully to the user, the user can access the passport by logging in to the system with the correct credentials. User with a passport can log in to the system, query the passport as well as apply for a VISA. Valid User applies for a VISA by choosing a country, VISA type and VISA duration. VISA application is submitted to the blockchain ledger and is queried by the VISA issuing agent.

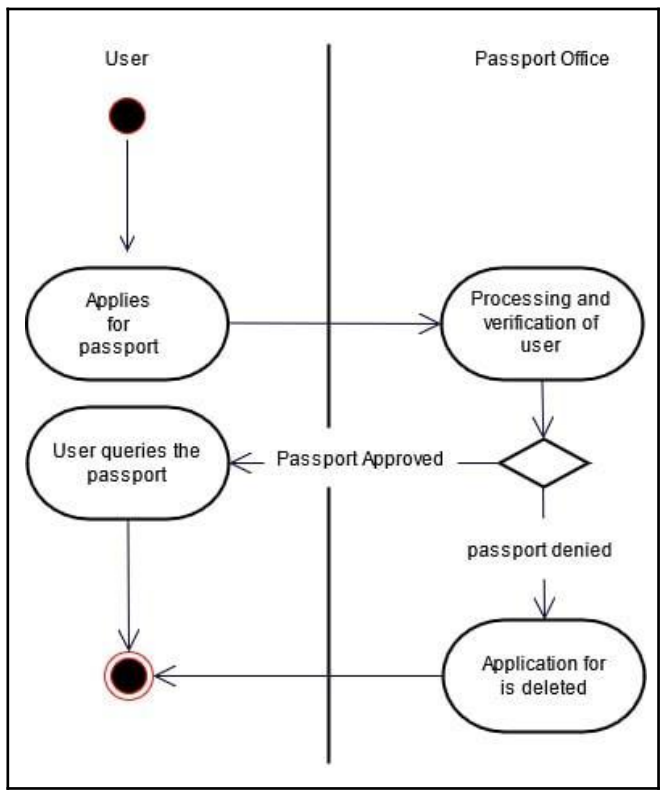

*Figure 4. Activity diagram for the user applying for passport*

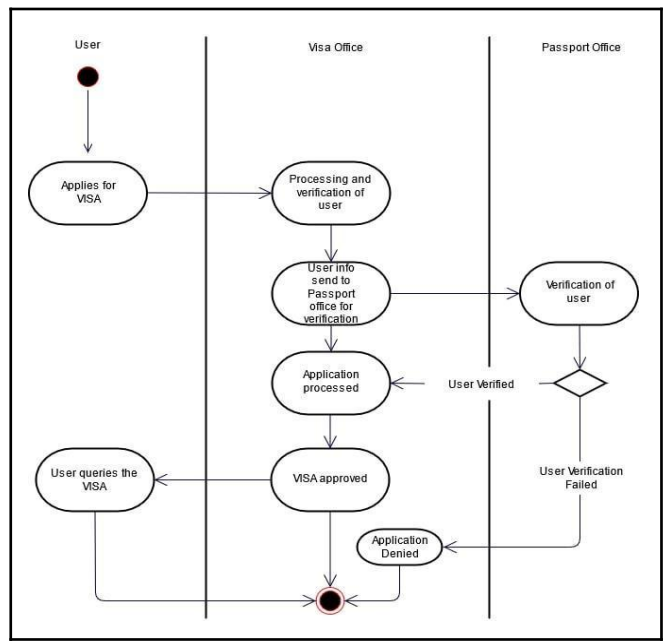

*Figure 5. Activity diagram for user applying for VISA*

VISA issuing agent needs correct credentials to log in to the system and query pending VISA requests. All the queries are verified by the Passport Office of the user's country. A VISA agent can accept or reject the passport based on the details entered by the user. VISA requests have two ways of getting rejected: the passport office of the user's country denies the verification or the VISA agent denies the VISA application.

## CONCLUSION

Existing solutions for maintaining assets like Passport and VISA based on blockchain cost unnecessary charges and work with the organisation sharing similar protocols and maintaining a shared chain network. This is not feasible in the current scenario, as the rules for immigration and VISA policy change from country to country. Hyperledger technology based on blockchain has a solution for both problems, cutting overhead cost every time transaction is done because it has no native tokens or cryptocurrency and solves latter problems by working with organisations even though different organisations use different protocols. It also supports multi-chain asset transfers.

## ACKNOWLEDGEMENT

We gratefully acknowledge Dr. Thippeswamy M. N., HoD, Department of Computer Science, Nitte Meenakshi Institute of Technology and our project guide Dr. Nalini N. Professor, Department of Computer Science, Nitte Meenakshi Institute of Technology for their valuable time and guidance.